# Stabilities and novel electronic structures of three carbon nitride bilayers

Wanxing Lin,[1] Shi-Dong Liang,[1] Chunshan He,[1] Wucheng Xie,[2] Haiying He,[3] Quanxiang Mai,[2]

Jiesen Li,[2,*] and D. X. Yao[1,†]

1. State Key Laboratory of Optoelectronic Materials and Technologies, School of Physics, Sun Yat-Sen University, Guangzhou, P. R. China
2. School of Environment and Chemical Engineering, Foshan University, Foshan, P. R. China
3. School of Materials Science and Energy Engineering, Foshan University, Foshan, P. R. China

E-mail: [*]2lgy@163.com, [†]yaodaox@mail.sysu.edu.cn

Three new novel phases of carbon nitride (CN) bilayer, which are named as $\alpha$-$C_2N_2$, $\beta$-$C_2N_2$ and $\gamma$-$C_4N_4$, respectively, have been predicted in this paper. All of them are consisted of two CN sheets connected by C-C covalent bonds. The phonon dispersions reveal that all these phases are dynamically stable, since no imaginary frequency is found for them. Transition path way between $\alpha$-$C_2N_2$ and $\beta$-$C_2N_2$ is investigated, which involves bond-breaking and bond-reforming between C and N. This conversion is difficult, since the activation energy barrier is found to be 1.90 eV per unit cell, high enough to prevent the transformation at room temperature. Electronic structures calculations show that they are all semiconductors with indirect band gap of 3.76 / 5.22 eV, 4.23 / 5.75 eV and 2.06 / 3.53 eV by PBE / HSE calculation, respectively. The $\beta$-$C_2N_2$ has the widest band gap among the three phases. From our results, the three new two-dimensional materials have potential applications in the electronics, semiconductors, optics and spintronics.

**Introduction**

Two-dimensional (2D) materials own much novel electronic and magnetic properties such as high mobility and optical characteristics[1][2]. The low-buckled honeycomb lattice that consists of silicon and germanium atoms, which are called silicene and germanene, have been predicted[3], one germanene exhibits quantum-spin Hall effect[4], the new materials were synthetized soon after its prediction[5][6]. The plumbene is a normal insulator at free state, then it can turn to be topological insulator by electron doping[7].

The 2D honeycomb monolayer consisting of nitrogen atoms named nitrogene has been proposed and the electronic properties have been deeply investigated[8][9]. The electronic properties of nitrogene with vacancy and adsorbed adatoms were also analyized[10]. Around the same time, another allotrope octagon-nitrogene has been proposed, and the stabilities and electronic structures have been systematically studied[11][12].

The number of 2D materials has been restricted due to the limited number of possible geometric structures, and new compound materials that contain more than one element are gaining increasing attentions. And the exploration of new 2D materials is in drastic need. Binary compounds based on two types of atoms may present new phenomenon comparing to their element counterparts[13]. The bulk structures formed by carbon and nitrogen atoms $C_3N_4$ have been predicted in recent years, the electronic and optic properties have been studied by first principle method[14]. Some other 2D carbon nitride materials can be used in 'post-silicon electronics'[15]. However, the carbon nitride bilayer have not proposed or studied until now.

We are trying to explore new 2D materials with novel properties by first principle study based on density functional theory (DFT). In this paper, we report a systematic study of stabilities and electronic structures of the previously unknown phases of 2D binary compounds that have the same formula $C_2N_2$, or $C_4N_4$, and we name them $\alpha$-$C_2N_2$ and $\beta$-$C_2N_2$, and $\gamma$-$C_4N_4$, respectively. Our results indicate that all of the three phases are dynamically stable, and all of them are indirect band gap insulators. Some other similar binary compounds with direct band gap have been predicted in our other studies[16].

## Computational Details

The calculations have been taken by the Vienna *Ab initio* Simulation Package (VASP) code[17] within Plane augmented wave and Perdew-Burke-Ernzerh potential[18]. Comparing to the PBE calculations, the hybrid function HSE06 calculations have been taken with a screened Coulomb potential[19][20]. The vacuum between two different bilayers is no less than 15 Å. Structures are relaxed until the net force on each ion is less than 0.0001 eV/Å. Brillouin zone was sampled with a $\Gamma$ centered grid of $20 \times 20 \times 1$ $k$ points. Phonon dispersions were calculated by combining the VASP with Phonopy[21]. During the calculation of force constants, a $4 \times 4 \times 1$ supercell has been taken for $\alpha$-$C_2N_2$ and $\beta$-$C_2N_2$, while $4 \times 6 \times 1$ supercell for $\gamma$-$C_4N_4$. The reaction path is calculated by the climbing image nudge elastic band methods (CINEB)[22][23][24], including five images between two kinds of structures. The spin-orbit coupling (SOC) is ignorable during the calculations.

## Results and Discussion
### Structure and Stabilities

The fully relaxed geometric structures of three carbon nitride bilayers are shown in figure 1. The $\alpha$-$C_2N_2$ and $\beta$-$C_2N_2$ bilayers have stable honeycomb structure, as shown in figure 1(a-b), therefore the Brillouin zone (BZ) is a regular hexagon (figure 2(d)). In figure 1 (a-b), the basis vectors are denoted as the red arrows, and the unit cells, which contain two carbon atoms and two nitrogen atoms, are denoted as green rhombs. Both of them have the same structure parameters with the in-plane lattice constant 2.35 Å, C-C bond length 1.62 Å, and C-N bond length 1.44 Å. However, they belong to the different point groups: $D_{3h}$ for $\alpha$-$C_2N_2$ and $D_{3d}$ for $\beta$-$C_2N_2$. The thickness of bilayer structure is 2.6 Å. The bond angle between the two C-N bonds is 108° 57'55", and the bond angle between the C-N bond and C-C bond is 109°58'21". Compared with the bond angle of a regular tetrahedral carbon with $sp^3$ hybridization, 109°28'16", the $\alpha$-$C_2N_2$ and $\beta$-$C_2N_2$ bilayers show slight deviation from the standard tetrahedral carbon structure. For the $\gamma$-phase, we find that $a_1$ is 2.35 Å, $a_2$ is 3.99 Å, the N-N bond length is 1.47 Å, the intralayer C-C bond length is 1.54 Å, and the C-N bond length is 1.45 Å, the thickness of the bilayer is 2.865 Å. The interlayer C-C bond length is 1.55 Å, which corresponds to the interlayer distance. Unlike the $\alpha$- and $\beta$-phases, the $\gamma$-$C_2N_2$ has the $D_{2h}$ symmetry. The basis vectors and unit cell are denoted in figure 1(c). Its unit cell contains four carbon atoms and four nitrogen atoms. The BZ has a rectangular shape, as shown in figure 2(e).

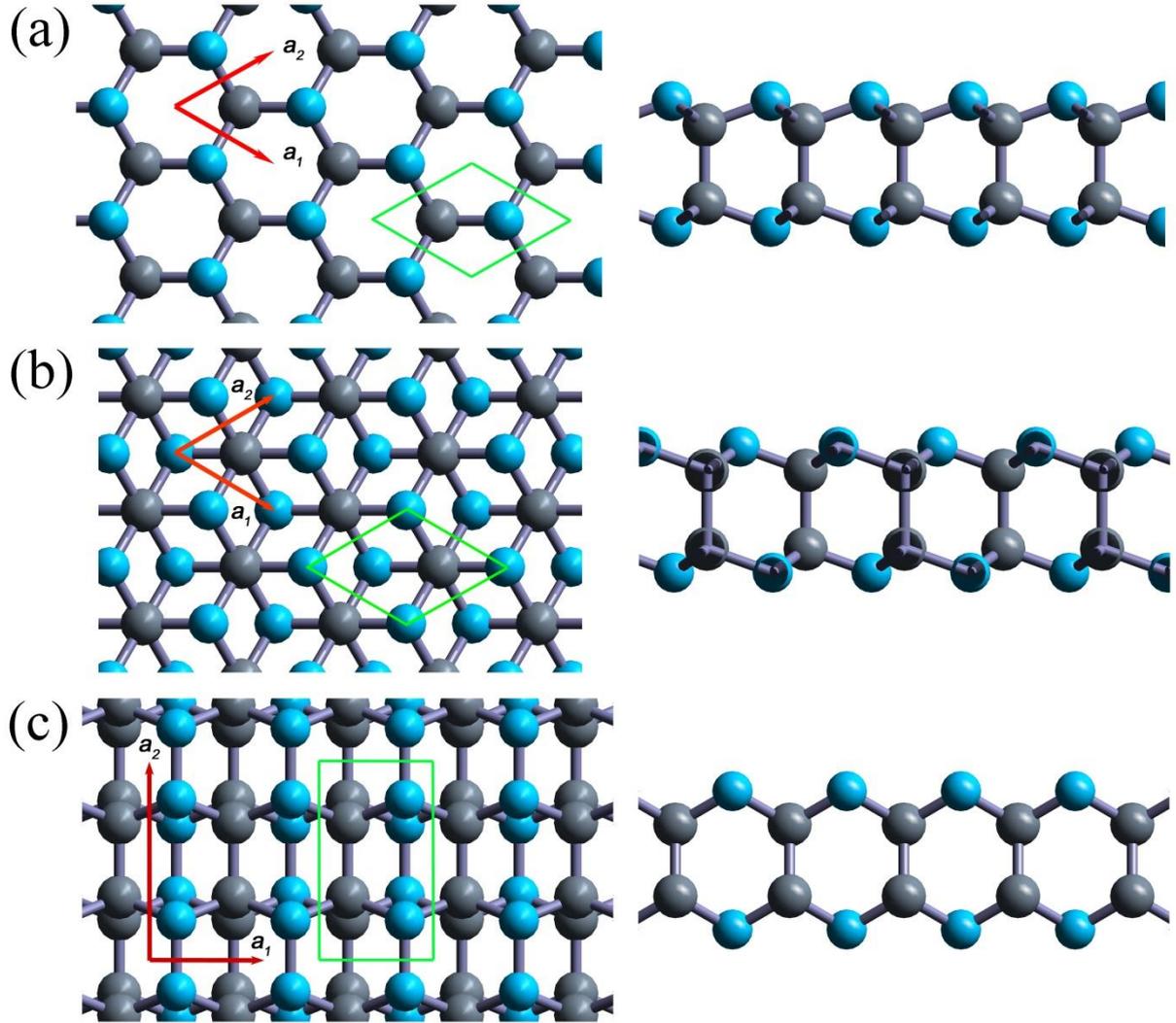

Figure 1. Geometric structures of three carbon nitride bilayers. Top and side views of (a) $\alpha$-$C_2N_2$, (b) $\beta$-$C_2N_2$ and (c) $\gamma$-$C_4N_4$, respectively. The grey and light blue spheres denote carbon and nitrogen atoms, respectively.

In order to explore the stabilities of three carbon nitride bilayers, we calculate their phonon dispersions shown in figure 2(a-c). There are no vibration modes with imaginary frequency along the high-symmetry (HS) lines in the whole BZ for all of the three phases, which suggests all of them are dynamically stable. Furthermore, no energy gap is found from the dispersions. The wave velocities have been fit from the phonon dispersions and listed in Table 1, which shows that both the $\alpha$-$C_2N_2$ and $\beta$-$C_2N_2$ bilayers are mostly isotropic, however, the $\gamma$-$C_4N_4$ bilayer exhibits great anisotropy from their different velocities from the $\Gamma$ point to other HS points.

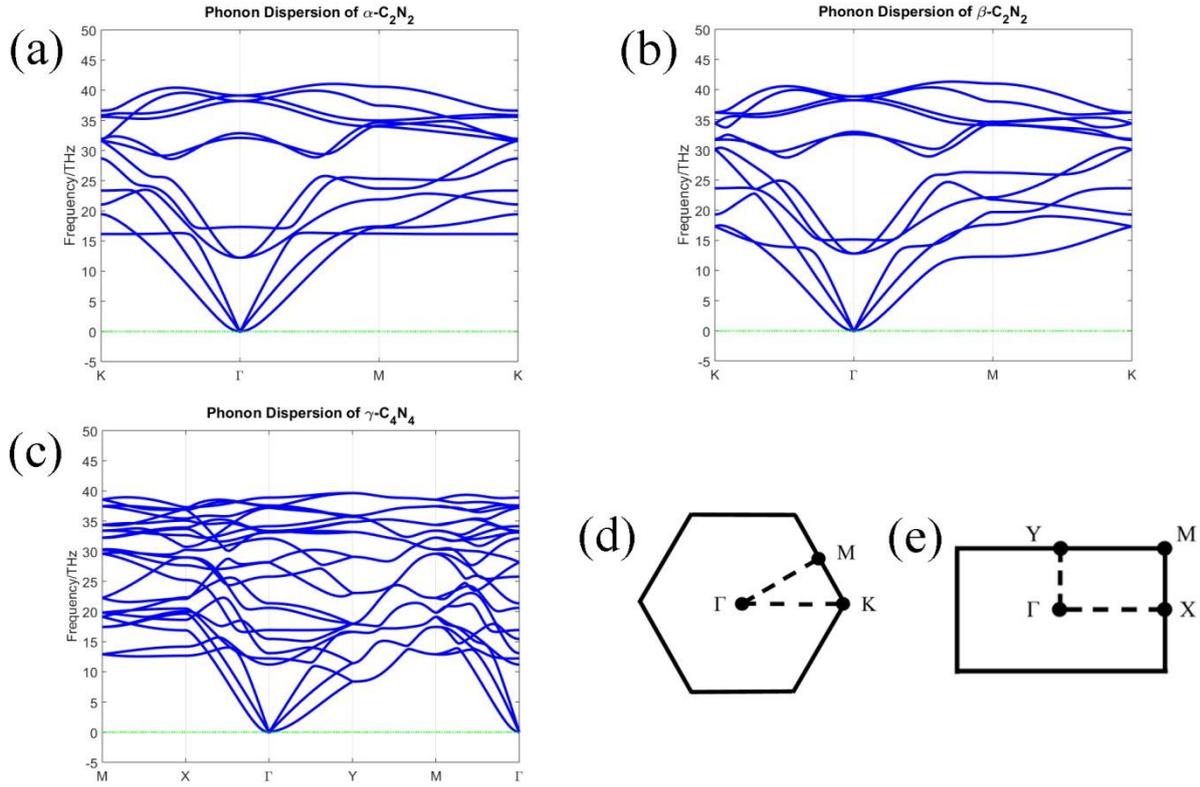

Figure 2. Phonon dispersions of (a) $\alpha$-$C_2N_2$, (b) $\beta$-$C_2N_2$, and (c) $\gamma$-$C_4N_4$ bilayers along the HS lines in the BZ, respectively. The BZs of (d) $\alpha$-$C_2N_2$, $\beta$-$C_2N_2$, and (e) $\gamma$-$C_4N_4$ bilayers.

Table 1. Wave Velocities of Acoustic Modes (m/s)

| Acoustic Modes | $\alpha$-$C_2N_2$ | | $\beta$-$C_2N_2$ | | $\gamma$-$C_4N_4$ | | |
|---|---|---|---|---|---|---|---|
| | $\Gamma \to M$ | $\Gamma \to K$ | $\Gamma \to M$ | $\Gamma \to K$ | $\Gamma \to X$ | $\Gamma \to Y$ | $\Gamma \to S$ |
| ZA | 609.7 | 639.0 | 645.9 | 641.9 | 65.84 | 874.1 | 694.6 |
| TA | 11630 | 11580 | 11560 | 11510 | 5351 | 5530 | 13810 |
| LA | 18570 | 18760 | 18350 | 18540 | 9417 | 8867 | 20290 |

Due to the similarity between the $\alpha$-$C_2N_2$ and $\beta$-$C_2N_2$, we have proposed a possible conversion path way between the two phases. From the CINEB calculation, and the energy change during the transformation is shown in figure 3(a). Since the CINEB can optimize one of the images to transition state, we study its vibration mode with imaginary frequency corresponding to the conversion path way, as shown in figure 3(b). Obviously, the transition state during the conversion involves the breaking and reforming of C-N bonds, and the transition has a comparatively high energy of 1.90 eV with respect to $\alpha$-$C_2N_2$ and $\beta$-$C_2N_2$ bilayers (figure 3(a)), which is equivalent to an activation energy of 183kJ/mol, a formidable energy barrier that can prevent reactions under normal conditions.

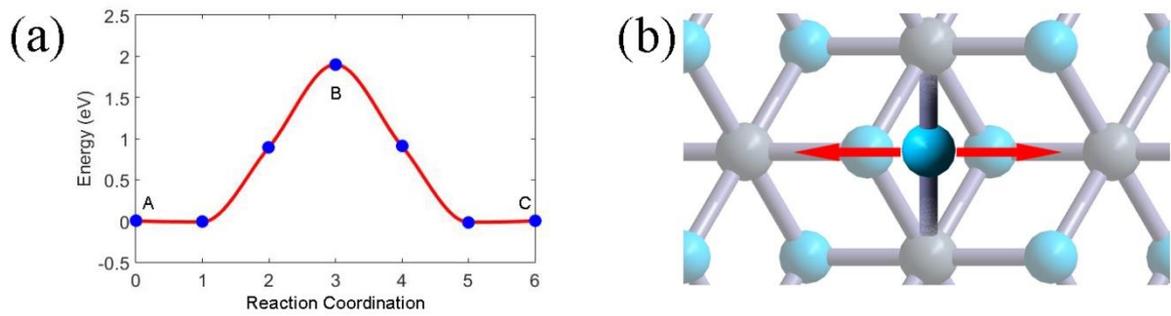

Figure 3. (a) Change of total energy of per unit cell during the transformation from $\alpha$-$C_2N_2$ to $\beta$-$C_2N_2$. The energy of $\alpha$-$C_2N_2$ is set to zero. Blue dots correspond to the inserted images during the CINEB calculation, the dot A corresponds to the $\alpha$-$C_2N_2$ bilayer, the dot B corresponds to the transition state , and the dot C corresponds to the $\beta$-$C_2N_2$ bilayer, respectively. Red curve is the spline interpolation. (b) The geometric structure of transition states. The vibration mode with imaginary frequency is presented by the double-ended arrow, corresponding to the conversion path along the transition.

**Electronic Structures**

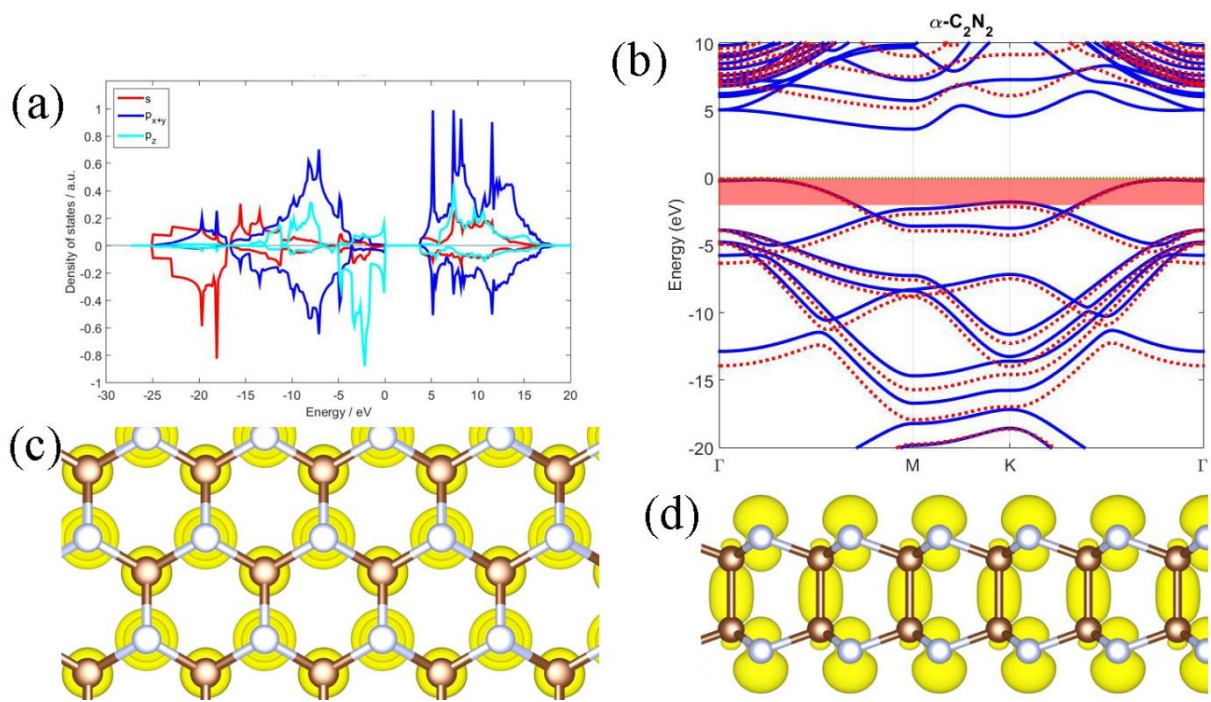

Figure 4. Electronic structures of the $\alpha$-$C_2N_2$ bilayer. (a) Density of states, the up and low part of the graph represents C and N components, respectively, and the red, blue and cyan lines represent the s, $p_{x+y}$, $p_z$ orbital projected densities of states, respectively. (b) The electronic band structures, blue solid/red dotted line represents the PBE / HSE calculation, the red shaded area indicates the region from the Fermi level to 2 eV below of PBE calculation. (c) Charge density of states in the red shaded area of (b), brown and silvery joints correspond to the carbon and nitrogen atoms, respectively. Fermi energy is set to zero.

In this subsection, the electronic structures of the three bilayers will be discussed. In order to investigate the orbital properties of the new phases, the projected density of states (PDOS) have been calculated. PDOS of orbitals of $\alpha$-$C_2N_2$ are shown in figure 4(a), with the up and low panel being the PDOS of carbon and nitrogen, respectively. As shown in figure 4(a), most of the low-energy states come from the s orbits of nitrogen atoms, while states from valance bands near the Fermi energy come from the $p_z$ orbit of nitrogen atoms. In the conducting bands, the $p_{x+y}$ orbit of both elements make the greater contribution than the s orbits and the $p_z$ orbit, and the carbon plays a major role in forming the conducting band. The PDOS of the $\beta$-$C_2N_2$ and $\gamma$-$C_4N_4$ are similar.

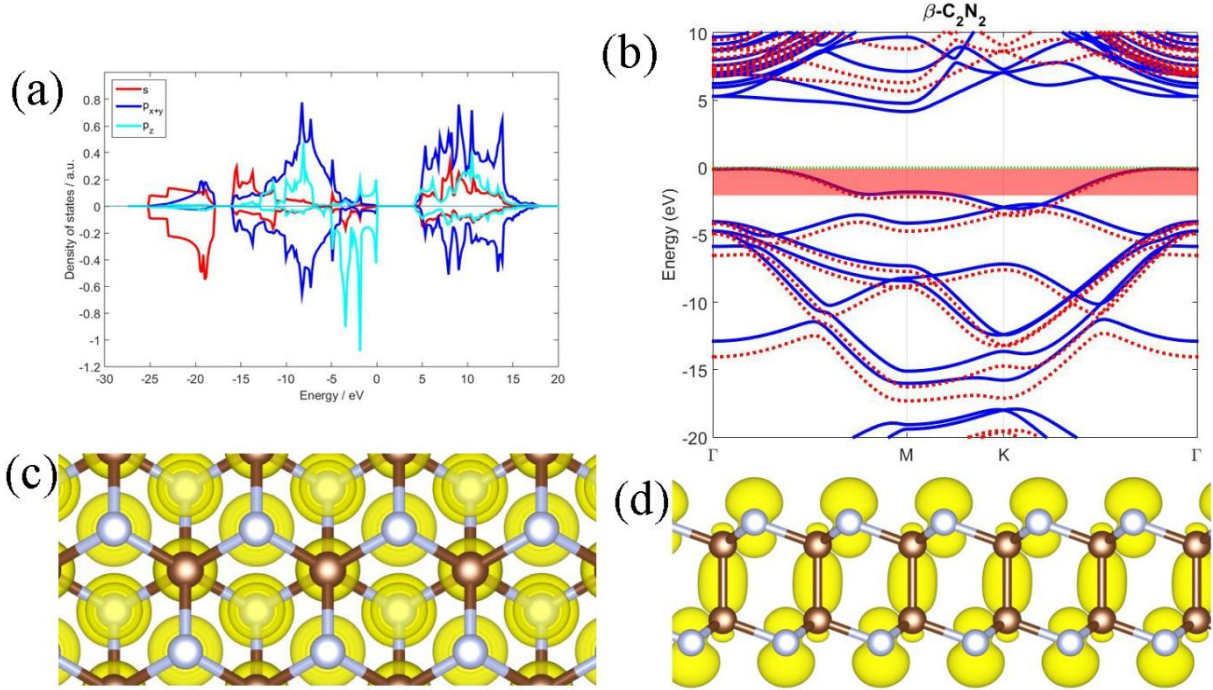

Figure 5. Electronic structures of the $\beta$-$C_2N_2$ bilayer. (a) Density of states, the up and low part of the graph represents C and N components, respectively, and the red, blue and cyan lines represent the s, $p_{x+y}$, $p_z$ orbital projected densities of states, respectively. (b) The electronic band structures, blue solid / red dotted line represents the PBE / HSE calculation, the red shaded area indicates the region from the Fermi level to 2 eV below of PBE calculation. (c) Charge density of states in the red shaded area of (b), brown and silvery joints correspond to the carbon and nitrogen atoms, respectively. Fermi energy is set to zero.

The electronic bands of the $\alpha$-$C_2N_2$ calculated by PBE and HSE are plot in figure 4(a). The results indicate that the $\alpha$-$C_2N_2$ is an indirect band gap semiconductor, with a 3.76 eV band gap by PBE calculation and a 5.22 eV band gap by HSE calculation, the PBE calculation underestimates the band gap of 1.46 eV than the HSE calculation. The PBE and HSE eigenvalue are nearly 'degenerate' near the Fermi level. The figure 5(b) shows the $\beta$-$C_2N_2$ is also a semiconductor with indirect band gap of 4.23 eV by PBE calculation and 5.75 eV by HSE calculation, the PBE calculation underestimates the band gap of 1.52 eV than the HSE calculation. The $\gamma$-$C_4N_4$ has an indirect band gap of 2.06 eV by PBE calculation, and the HSE calculation gives a wider one with 3.53 eV, which is 1.47 eV wider than the PBE band from figure 6(b). The band gap of $\alpha$-$C_2N_2$ is near the band gap of nitrogene[8][9], and the $\beta$-$C_2N_2$ has the widest band gap, while the $\gamma$-$C_4N_4$ has the narrowest one among of them. Furthermore, their conducting band minimum (CBM) of three carbon nitride bilayers are all located at the M point, then the valence band maximum (VBM) of the $\alpha$-$C_2N_2$ and $\beta$-$C_2N_2$ bilayers are located along the K-$\Gamma$ line, while the VBM of the $\gamma$-$C_4N_4$ is located at the $\Gamma$ point. The $\gamma$-$C_4N_4$ bilayer can be controlled to a direct band gap semiconductor by the uniform tension or electric field since the VBM is at the Γ point.

In order to clarify the bonding mechanism of the new structures, the partial charge densities (PCD) are calculated. The charge density in the area from the Fermi level to 2 eV below of $\alpha$-$C_2N_2$ is included in figure 4(c-d), the charge density around nitrogen atoms exhibits sphere outside out the surface, while the charge density exhibits ellipsoid around the C-C bond. Most states are located around the nitrogen atoms, which are mostly formed by the $p_z$ orbit, because the nitrogen atom owns one more electron than the carbon atom in the outer sphere. The other two phases exhibit similar phenomenon, see figure 5(c-d) and figure 6(c-d). The nitrogen atoms and carbon atoms in the three structures prefer the $sp^3$ hybridization by combining the geometry structures with orbit characteristics.

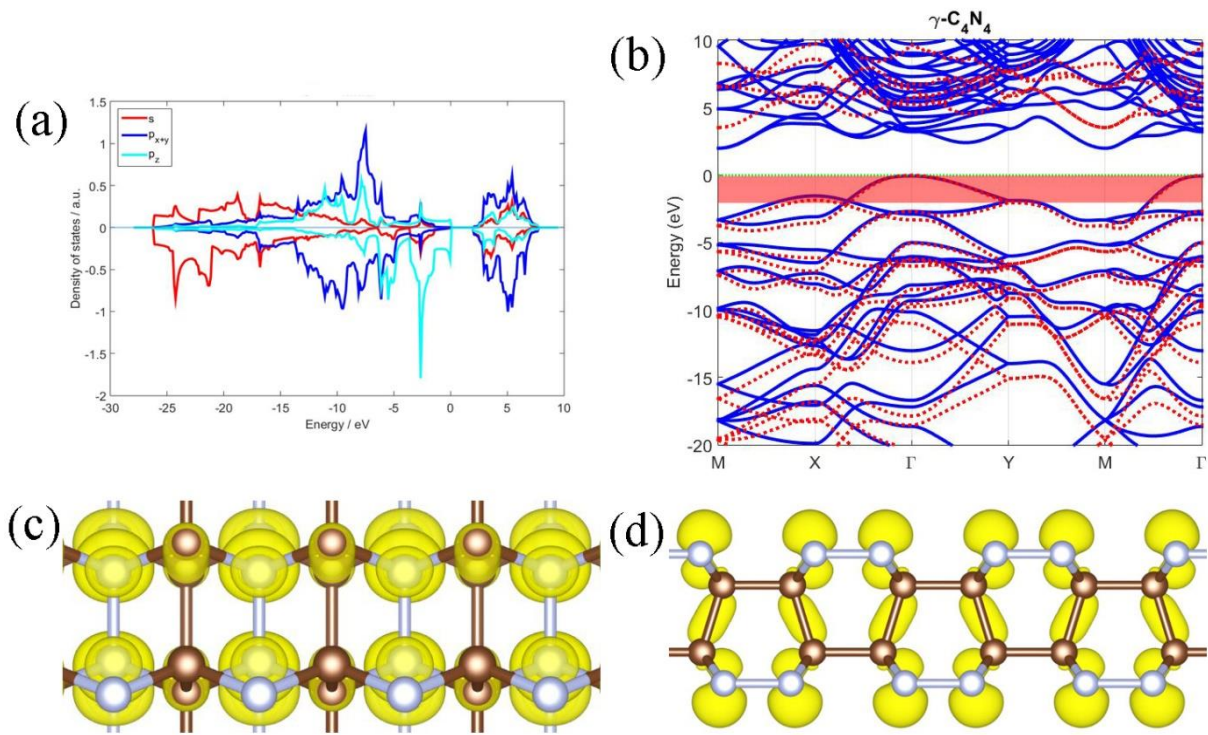

Figure 6. Electronic structures of the $\gamma$-$C_2N_2$ bilayer. (a) Density of states, the up and low part of the graph represents C and N components, respectively, and the red, blue and cyan lines represent the s, $p_{x+y}$, $p_z$ orbital projected densities of states, respectively. (b) The electronic band structures, blue solid / red dotted line represents the PBE / HSE calculation, the red shaded area indicates the region from the Fermi level to 2 eV below of PBE calculation. (c) Charge density of states in the red shaded area of (b), brown and silvery joints correspond to the carbon and nitrogen atoms, respectively. Fermi energy is set to zero.

## Conclusions

Three novel phases of carbon nitride bilayers have been predicted by density function theory method. Their phonon dispersions and electronic structures have been calculated. Our results that all of them are stable, while the conversion between the $\alpha$-$C_2N_2$ and $\beta$-$C_2N_2$ bilayers is found to be difficult due to the high energy barrier. The electronic bands calculated both by PBE and HSE method indicate that the three bilayer structures are all indirect semiconductors. Among them, the $\beta$-$C_2N_2$ bilayer has the widest band gap. The $\gamma$-$C_2N_2$ bilayer perhaps can become a direct semiconductor by using the strain or electric field. These novel bilayer materials may be be used in many fields such as electronics, semiconductors, spintronics, batteries, and supercapacitors[25].

**Acknowledgments**

W. L. and D. X. Y. are supported by National Key R&D Program of China 2017YFA0206203, NSFC-11574404, NSFC-11275279, NSFG-2015A030313176, Special Program for Applied Research on Super Computation of NSFC-Guangdong Joint Fund, Leading Talent Program of Guangdong Special Projects. J. L. is supported by the Opening Project of Guangdong High Performance Computing Society (2017060103), High-Level Talent Start-Up Research Project of Foshan University (Gg040904). S. D. L. is supported by the Natural Science Foundation of Guangdong Province (No. 2016A030313313). W. X is supported by the Natural Science Foundation of Guangdong Province （2017A030310577）. All calculations of this work were performed on Tianhe-2 supercomputer with the help of engineers from National Supercomputer Center in Guangzhou and Paratera.